\documentclass[prx,10pt,twocolumn,groupedaddress,floatfix, showpacs,nofootinbib]{revtex4-2}

\usepackage[sc,osf]{mathpazo}
\usepackage[scaled=0.86]{berasans}  
\usepackage[colorlinks=true, allcolors=blue, urlcolor=blue]{hyperref}  
\usepackage{graphicx} 
\usepackage{amsmath,mathtools,amssymb,amsthm,bm,amsfonts,mathrsfs,bbm} 

\usepackage{xspace}  
\usepackage{pgfplots}
\usepackage{xcolor,colortbl}
\usepackage{array}
\usepackage{bigstrut}
\usepackage{mathrsfs}
\usepackage{dsfont}
\usepackage{quantikz}
\usepackage{multirow}
\usepackage{tikz}
\usepackage{quantikz}

\usepackage{caption}
\usepackage{subcaption}
\usepackage{ragged2e}
\DeclareCaptionJustification{justified}{\justifying}
\usepackage{tabularx}
\newcolumntype{C}{>{\centering\arraybackslash}X}

\usepackage{lipsum}

\usepackage{verbatim}

\usepackage{comment}

\newcommand{\N}{\mathbb{N}}

\newcommand{\tr}{\text{tr}}

\newcommand{\bracket}[3]{\langle #1|#2|#3 \rangle}

\newcommand{\be}{\begin{equation}}
\newcommand{\ee}{\end{equation}}
\newcommand{\bea}{\begin{eqnarray}}
\newcommand{\eea}{\end{eqnarray}}
\newcommand{\bes}{\begin{equation*}}
\newcommand{\ees}{\end{equation*}}
\newcommand{\beas}{\begin{eqnarray*}}
\newcommand{\eeas}{\end{eqnarray*}}

\newcommand{\I}{\mathds{1}}

\def\M{\widetilde{M}}
\def\N{\widetilde{N}}

\def\tr{\mathrm{tr}}

\begin{document}



\title{ Static Quantum Errors and Purification}

\author{Jaemin Kim}
\email{woals6584@kaist.ac.kr} 

\author{Seungchan Seo}
\email{ichi9505@kaist.ac.kr}

\author{Jiyoung Yun}
\email{jiyoungyun@kaist.ac.kr}

\author{Joonwoo Bae}
\email{joonwoo.bae@kaist.ac.kr}

\affiliation{School of Electrical Engineering, Korea Advanced Institute of Science and Technology (KAIST), $291$ Daehak-ro, Yuseong-gu, Daejeon $34141$, Republic of Korea }


\begin{abstract}
State preparation that initializes quantum systems in a fiducial state and measurements to read outcomes after the evolution of quantum states, both essential elements in quantum information processing in general, may contain noise from which errors, in particular, referred to as {\it static errors}, may appear even with noise-free evolution. In this work, we consider noisy resources such as faulty manufacturing and improper maintenance of systems by which static errors in state preparation and measurement are inevitable. 
We show how to suppress static errors and purify noiseless SPAM by repeatedly applying noisy resources. We present the purification protocol for noisy initialization and noisy measurements and verify that a few qubits are immediately cost-effective to suppress error rates up to $10^{-3}$. We also demonstrate the purification protocol in a realistic scenario. The results are readily feasible with current quantum technologies.

\end{abstract}

\maketitle


{\it Introduction.} Errors persisting in a quantum information task even when a quantum evolution is noiseless verify the existence of imperfections in state preparation and measurement (SPAM), see e.g., \cite{PhysRevA.100.052315, 9142431}. They are referred to as {\it static errors}, which can be generic from faulty manufacturing of quantum devices or temporal due to improper maintenance, see e.g. \cite{Zwerver:2022aa, Pauka:2021aa}; nonetheless, such errors are not fixed by quantum error correcting codes which deal with gate errors in a quantum circuit. In the other way around, quantum error correction needs noiseless SPAM to detect errors, called syndrome measurements, in order to find how to fix errors in a circuit, see, e.g., \cite{PhysRevA.52.R2493, 681315, doi:10.1098/rspa.1996.0136, Devitt_2013}. 
 
In practical applications, static quantum errors are naturally present in two ways, as follows. On the one hand, the SPAM of multiple qubits enables quantum error correction. As the most straightforward instance, a single qubit controlled by the two qubits via a controlled-NOT gate (CNOT), a fundamental building block in quantum error correction, can detect a parity of two qubits in a logical state, called syndrome measurements. The measurement outcomes can be trusted, given noiseless SPAM for the parity check. While both the SPAM and the CNOT gate can be noisy, little is known about the effect of SPAM errors, whereas much has been devoted to dealing with gate errors and establishing fault-tolerant quantum circuits \cite{doi:10.1137/S0097539799359385, 548464, doi:10.1126/science.279.5349.342}. The static errors may not directly propagate over a circuit, contrasting gate errors, but may mislead syndrome measurements, thus signifying incorrect operations to fix errors in error correction. Henceforth, static quantum errors affect quantum error correction and, therefore, may indirectly propagate over a circuit. 

\begin{figure*}[t]
    \centering
    \includegraphics[width=1.0\textwidth]{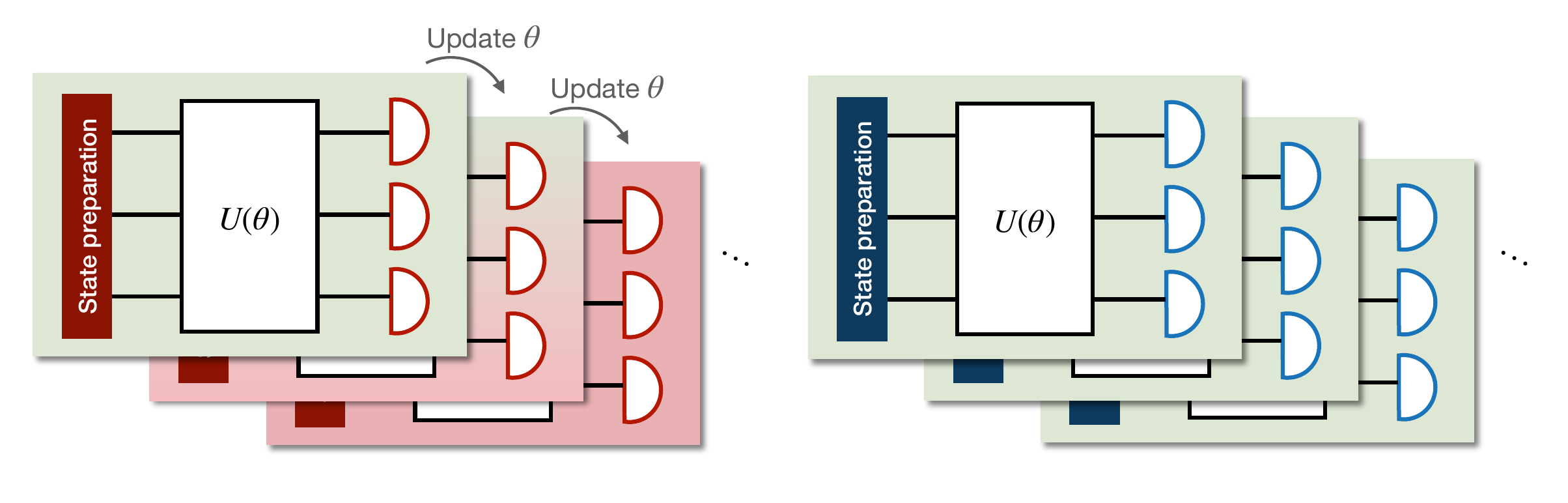}
    \caption{  One of the central applications of NISQ technologies to computing tasks such as optimization or training with the framework of variational quantum algorithms is to iterate short-depth quantum circuits via repetitions of SPAM. Parameters $\theta$ are trained, where static errors from noisy SPAM dominate the computational tasks.}
  \label{fig:vqe}
\end{figure*}

On the other hand, in the era of noisy-intermediate-scale quantum (NISQ) technologies characterized by SPAM with tens of qubits but noisy operations, SPAM may be considered a free task that one can repeatedly apply without a cost. CNOT gates creating quantum interactions and the possibility of errors appearing in doing so are resources that need to be minimized since quantum error correction is yet lacking. Consequently, variational quantum algorithms are a well-known instance of NISQ applications for diverse purposes, such as optimization or training, in such a way that they iterate short-depth circuits so as not to face accumulating errors too fast, where realizations of the SPAM are therefore essential, see Fig. \ref{fig:vqe}, see also \cite{RevModPhys.94.015004, TILLY20221}. In the era of NISQ technologies, static quantum errors amount to two-qubit gates, such as a CNOT gate, and thus may correspond to a source that generates noise, preventing applications of NISQ technologies from working properly as designed.

From the view taken toward a fundamental understanding, we may distinguish static quantum errors from gate errors; obviously, one correction technique does not work for the other. Correction of gate errors, which aims to manipulate quantum states in a noiseless manner, actively explores where and what errors occur and finds a prescription. As mentioned, a parity check for diagnosing gate errors is also possible with quantum gates, such as a CNOT gate, and immediately asks for a sufficiently small error rate to reach the level of a fault-tolerant quantum circuit \cite{doi:10.1137/S0097539799359385, 548464, doi:10.1126/science.279.5349.342}. 

Static errors are not necessarily due to a failure in the control of quantum systems; they can happen in or even after measurements, e.g., because of instrument defects in chip-scale manufacturers, errors in reading, or random errors caused by unpredictable reasons. The inability to set a fiducial setting leads to static errors. Noiseless SPAM aims at a fiducial setting, initialization of quantum systems in a state, e.g., $|0\rangle\otimes\cdots \otimes |0\rangle$, and a measurement is in the computational basis, e.g., $\{ |0\rangle, |1\rangle \}$ \cite{nielsen_chuang_2010}. The scenario shares it common with entanglement distillation to put a fiducial setting in advance, where a particular type of entangled states, maximally entangled ones, is desired \cite{PhysRevLett.76.722, Bennett_1996, PhysRevLett.81.5932}. 

Hence, one may tackle the problem of static quantum errors similarly to entanglement distillation, which exploits a two-way key distillation protocol first proposed in a protocol for information-theoretic secret key agreement \cite{256484, 748999}. The strategy may follow from the lesson learned by connecting cryptographic privacy and quantumness on an equal footing \cite{PhysRevA.65.032321, oppenheim2008classical}. Remarkably, it works to devise a protocol for purifying a noisy measurement with noiseless state preparation \cite{kimyunbae2024}. It is also worth pointing out that, similarly to the two-way cryptographic protocol tolerating a higher error rate \cite{256484, 748999}, see also the device-independent setting \cite{PhysRevLett.124.020502, Hahn:2022aa}, its applications to quantum distillation tasks lead to quantum protocol tolerating much higher error rates; all two-qubit entangled states can be distilled \cite{PhysRevLett.76.722, Bennett_1996}, and all two-qubit entangled states can be used for quantum cryptographic protocols to establish secret bits \cite{PhysRevLett.91.167901}, see also for general considerations \cite{PhysRevA.75.012334, PhysRevA.72.032301}. The purification protocol for noisy measurements works unless an error rate is completely unbiased \cite{kimyunbae2024}. The results clarify a distinction in the approaches to and results from static and gate errors and, interestingly, strengthen connections between cryptographic privacy and quantumness.

The goal is to suppress static errors, the consequence of both states and measurements, at full generality. Thus, we identify the problem by asking about the possibility of purifying noisy SPAM with noisy SPAM. The problem is of practical significance and fundamental interest since it asks if noisy SPAM itself suffices to cleanse noisy SPAM. In quantum error correction and entanglement distillation, noiseless tasks such as measurements and local operations are essential to cleanse errors in quantum states; otherwise, they do not work. In addition, static errors cannot be used to diagnose their origin, preparation or measurement. 


In this work, we affirmatively answer the question addressed above: noisy SPAM {\it per se} can purify both a noisy state preparation and a noisy measurement, respectively. That is, none of the following, noiseless SPAM or verifying error rates of preparation and measurements, is necessary to execute cleansing noisy SPAM. We introduce the purification protocol inspired by the advantage distillation protocol from the information-theoretic secret key agreement and show the purification of noisy SPAM. We analyze resources for the purification and find that a few additional qubits are immediately cost-effective, reaching static error rates up to $10^{-3}$. We then present a demonstration of the purification of noisy SPAM with a few additional qubits. The results are readily feasible with NISQ technologies and present a pathway to exploit noisy quantum instruments towards quantum-enabled information tasks.  

\begin{figure*}[t]
    \centering
    \includegraphics[width=1.0\textwidth]{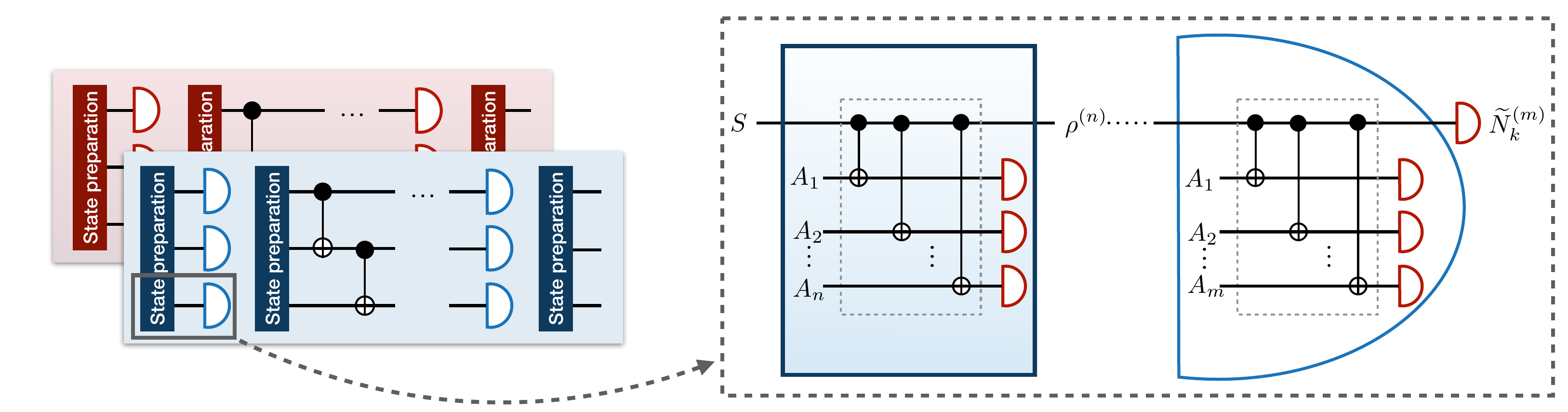}
    \caption{  Noisy initialization and noisy measurements are a building block in quantum information processing where quantum error correction is lacking. Noisy SPAM is purified by noisy resources of additional qubits. A collective CNOT in Eq. (\ref{eq:cv}) applies to noisy systems $SA_{1}\cdots A_n$. A resulting state $\rho^{(n)}$ accepted once the outcomes of noisy measurements on noisy qubits $A_1\cdots A_n$ are all zeros $0^n$ has a fidelity in Eq. (\ref{eq:f}) arbitrarily close to $1$ as $n$ increases. A noisy measurement on a system $S$ is purified with noisy SPAM with $m$ target qubits; a purified POVM element giving an outcome $k$ is denoted by $\N_{k}^{(m)}$, in which the noise fraction converges to $0$, see Eq. (\ref{eq:pm}). }
  \label{fig:scenario}
\end{figure*}

\subsection*{Static quantum errors}

Let us begin with the characterization of static errors. We recall that as mentioned, noiseless SPAM aims at a fiducial setting, a noiseless preparation sets a fiducial state, denoted by $|0\rangle$, at initialization and measurement in the computational basis, $\{M_0,M_1\}$ where $M_i = |i\rangle\langle i |$ for $i=0,1$, corresponds to a noiseless setting. We can describe {\it all} static quantum errors that take all types of imperfections into account as noisy SPAM, see Fig. \ref{fig:scenario}. Then, noisy SPAM is described by a probabilistic mixture of a fiducial setting and other possible ones. For instance, an instrumental defect in a detector leading to a fault in reporting, e.g., an outcome $k$ having a probability $p(k)$ reported as $k\oplus1$ with a probability $q$, due to some unpredictable reasons, shows a noisy probability of getting an outcome $k$, 
\bea
\widetilde{p}(k) = (1-q)p(k) + qp(\bar{k}). \nonumber
\eea
where $\bar{k} =k \oplus 1$ with a bitwise addition$\oplus$ and $q\in[0,1/2)$ a noise fraction. The noisy probability above can be equivalently modeled by a mixture of POVM elements, 
\bea
\widetilde M_{k} = (1-q) M_{k} + q M_{\bar{k}},\label{eq:nm}
\eea
without a fault in reporting, i.e., $\widetilde{p}(k) = \tr[\sigma \M_k]$ for a state $\sigma$. Then, a noise fraction defined in the following,
\bea
q =  \langle \bar{k}| \M_k | \bar{k}\rangle \label{eq:nf}
\eea
quantifies noise existing in a POVM element. 

As for state preparation, we write a noisy preparation by $\rho$, a state deviated from the desired one, 
\[
\rho =
\begin{bmatrix}
\rho_{00} & \rho_{01} \\
\rho_{10} & \rho_{11}
\end{bmatrix}
\]
and the preparation may be quantified by a fidelity 
\bea
f = \langle 0| \rho |0\rangle. \label{eq:f}
\eea 
We assume that a preparation is not unbiased and has $f>1/2$ throughout; otherwise, a NOT operation with a Pauli-$X$ gate is applied. A purification corresponds to a transformation from a state having $f<1$ to a desired one with $f=1$.

\subsection*{Purification of noisy state preparation}

We here show that noisy SPAM above producing static errors can be purified by repeatedly applying noisy SPAM themselves. Let us begin with the purification of noisy preparation. The protocol starts by generating $(n+1)$ copies of a noisy state $\rho$, feasible with a given preparation setting, where a fidelity above $0.9$ can be achieved with NISQ technologies. A collective CNOT gate for $(n+1)$ qubits is defined as follows,
\bea
V_{n}^{SA_1\cdots A_{n }} = |0\rangle \langle 0|^S \otimes \I^{\otimes  n  } + |1 \rangle \langle 1|^S \otimes X^{\otimes  n  } \label{eq:cv}
\eea
where the control qubit ($S$) is in the first register, $n$ target qubits ($A_i$) for $i=1,\cdots, n$ are in the next $n$ registers, and $X$ denotes a Pauli-$X$ gate. Note also that $V_n$ may be realized by a sequence of CNOT gates, $V_{n}^{SA_1\cdots A_{n }} =  \Pi_{i=1}^n V_{2}^{SA_i}$, see also Fig. \ref{fig:scenario}. 

After an application of a collective CNOT gate, measurements are performed on $n$ target qubits with noisy POVM elements in Eq. (\ref{eq:nm}). When all of the $n$ outcomes on target qubits are zeros, i.e., $0^{n}$, a resulting state in the control register is accepted, denoted by $\rho^{(n)}$, which may quantified by the fidelity $f^{(n)} = \langle 0| \rho^{(n)} |0\rangle $. In what follows, we show the purification, i.e., the fidelity is arbitrarily close to $1$ as $n$ increases.

We write by $R^{(n)}$ the unnormalized state accepted after an interaction $V_n$ over a control and $n$ target qubits,  
\bea
R^{(n)} =  \begin{bmatrix}
R^{(n)}_{00} & R^{(n)}_{01} \\
R^{(n)}_{10} & R^{(n)}_{11} 
\end{bmatrix}. \label{eq:uns}
\eea
Note that the probability that a control qubit is accepted is given by $p_{succ}^{(n)} = \tr R^{(n)}$, and the fidelity is thus as follows, 
\bea
f^{(n)} =  \frac{1}{p_{succ}^{(n)}}  R_{00}^{(n)}. \label{eq:fn}
\eea
Let us exploit a decomposition of a collective CNOT gate in Eq. (\ref{eq:cv}),
\bea
V_{n+1}^{S A_1\cdots A_{n+1} } = V_{2}^{SA_{n+1}} V_{n}^{S A_1\cdots A_{n}  }. \label{eq:decom}
\eea
to derive a recurrence relation for a resulting unnormalized state accepted by measurements on target qubits,
\bea
R^{(n+1)} = \tr_{A_{n+1}} [ V_{2} (~ R^{(n)} \otimes  \rho^{A_{n+1}} ) ~ V_{2}^{\dagger} \widetilde{M}_{0}^{A_{n+1}}] \label{eq:res}
\eea
where $V_2$ acts on systems $S$ and  $A_{n+1}$. The elements also satisfy the following relations, 
\bea
R^{(n+1)}_{00} = \alpha R^{(n)}_{00} ~~\mathrm{and}~~ R^{(n+1)}_{11} = (1-\alpha)R^{(n)}_{11}~~~~~~~\label{eq:r}
\eea
where $\alpha = (1-q)f + q(1-f)$. It is straightforward to see that $1-2\alpha<0$, from which $\alpha>1/2$. It follows that
\bea
\frac{R^{(n+1)}_{11}}{R^{(n+1)}_{00}} = \left( \frac{1 - \alpha}{ \alpha} \right) \frac{R^{(n )}_{11}}{R^{(n )}_{00}} =\cdots=\left( \frac{1 - \alpha}{ \alpha} \right)^{n+1} \frac{R^{(0 )}_{11}}{R^{(0 )}_{00}} ~~~~~~\label{eq:re}
\eea
which converges to $0$ as $n$ increases, since $\alpha>1/2$. Thus, we have shown that the fidelity in Eq. (\ref{eq:fn}) is arbitrarily close to $1$ as $n$ tends to be large. Note that from the recurrence relation in Eq. (\ref{eq:r}), the success probability is also computed, $p_{succ}^{(n)} =  \alpha^n \rho_{00}  + (1-\alpha)^n \rho_{11}$. As a realistic instance, for balanced error rates $1-f = q = 0.05$, it holds $f^{(n)} > 0.999$ for $n \geq 2$. The success probability for $n=2$ is given by $0.7785$. Thus, a few additional qubits are immediately cost-effective and a success probability is high enough over $2/3$.

\subsection*{Purification of a noisy POVM element}

We next present the purification of a noisy measurement with noisy SPAM by suppressing a noise fraction $q$ arbitrarily close $0$. We begin with noisy POVM elements in Eq. (\ref{eq:nm}) and write by $ \widetilde{N}_{k}^{(m)}$ a resulting POVM element giving an outcome $k\in\{0,1 \}$ on a system qubit by sacrificing $m$ additional target qubits, see also Eq. (\ref{eq:nf}). A purified POVM element may be quantified as follows
\bea
q^{(m)} = \langle \bar{k}| \widetilde{N}_{k}^{(m)} | \bar{k}\rangle, \label{eq:nfm}
\eea
which we show converges to $0$ as $m$ tends to be large.

The protocol with noisy SPAM, see also Fig. \ref{fig:scenario}, begins by  applying a collective CNOT $V_m$ in Eq. (\ref{eq:cv}) to a qubit state $\sigma$ in a system and $m$ noisy target qubits prepared in noisy states $\rho^{\otimes m}$. Noisy measurements in Eq. (\ref{eq:nm}) are performed on the $m$ target qubits, which are accepted whenever the outcomes are identical, either $0^m$ or $1^m$. Then, a measurement performed on a system is accepted when an outcome $k$ is equal to the other results of $m$ qubits. 

A purified POVM element with $m$ target qubits can be described as follows. The probability of having identical outcomes  $k^{m+1}$ for $k\in\{0,1\}$ is given by, for a state $\sigma$ on a system, 
\bea
\tr_{SA_1\cdots A_m} [V_m ~\sigma^S \otimes \rho^{\otimes m} ~V_{m}^{\dagger} ~ \M_{k}^{\otimes m+1} ]. \nonumber
\eea
From above, we write by $p^{(m)}(k)$ a probability of having an outcome $k$ on a system once noisy measurements on a system and $m$ target qubits give identical outcomes, 
\bea
p^{(m)} (k) & = & \frac{1}{p_{succ}^{(m+1)}} \tr [ \sigma \M_{k}^{(m)}] ~~\mathrm{where} \label{eq:pum} \\
\M_{k}^{(m)} & = & \tr_{ A_1\cdots A_m} [    \rho^{\otimes m} V_{m}^{\dagger} ~ \M_{k}^S \otimes  \bigotimes_{i=1}^m \M_{k}^{A_i} ~ V_m ]  \nonumber
\eea
The decomposition in Eq. (\ref{eq:decom}) applies to derive a recurrence relation for a POVM element $\M_{k}^{(m)}$
\bea
\M_{k}^{(m+1)} & = & \tr_{ A_{m+1} } [ \rho^{A_{m+1}}    V_{2}^{  \dagger}   ~    \M_{k}^{(m) } \otimes  \M_{k}^{ A_{m+1} } ~ V_{2}^{ } ] \label{eq:rem}
\eea
where $V_2$ acts on systems $S$ and $A_{m+1}$. It also follows that
\bea
\bracket{k}{ \M_k^{(m+1)}}{k}  & = & \alpha \bracket{ k }{ \M_k^{(m)}}{ k } ~\mathrm{and}~  \nonumber \\
\bracket{\Bar{k}}{\M_k^{(m+1)}}{\Bar{k}} & = & (1-\alpha)  \bracket{\Bar{k}}{ \M_k^{(m  )}}{\bar{k}}, \nonumber
\eea
where $\alpha = f(1-q) + (1-f)q$. It may be straightforward to find that
\bea
\M_{k}^{(m)} = \alpha^m(1-q) |k\rangle\langle k| +  (1-\alpha)^m q |\bar{k}\rangle \langle \bar{k}|. \nonumber
\eea
Then, from Eq. (\ref{eq:pum}), a purified POVM element $\N_{k}^{(m)}$ corresponds to, 
\bea
\N_{k}^{(m)} =\frac{1}{p_{succ}^{(m+1)} } \M_{k}^{(m)} \label{eq:pm}
\eea
where $p_{succ}^{(m+1)} =  \alpha^{m} (1-q) + (1-\alpha)^{m} q$. One finds that a noise fraction converges to zero, see also Eq. (\ref{eq:nfm})
\bea
q^{(m)} = \frac{ (1-\alpha)^m q }{ \alpha^{m} (1-q) + (1-\alpha)^{m} q }~~  \xrightarrow{ m \rightarrow \infty }~~ 0. \nonumber
\eea
It is thus shown that the protocol can purify noisy POVM elements.

\begin{figure*}[t]
    \centering
    \includegraphics[width=1\textwidth]{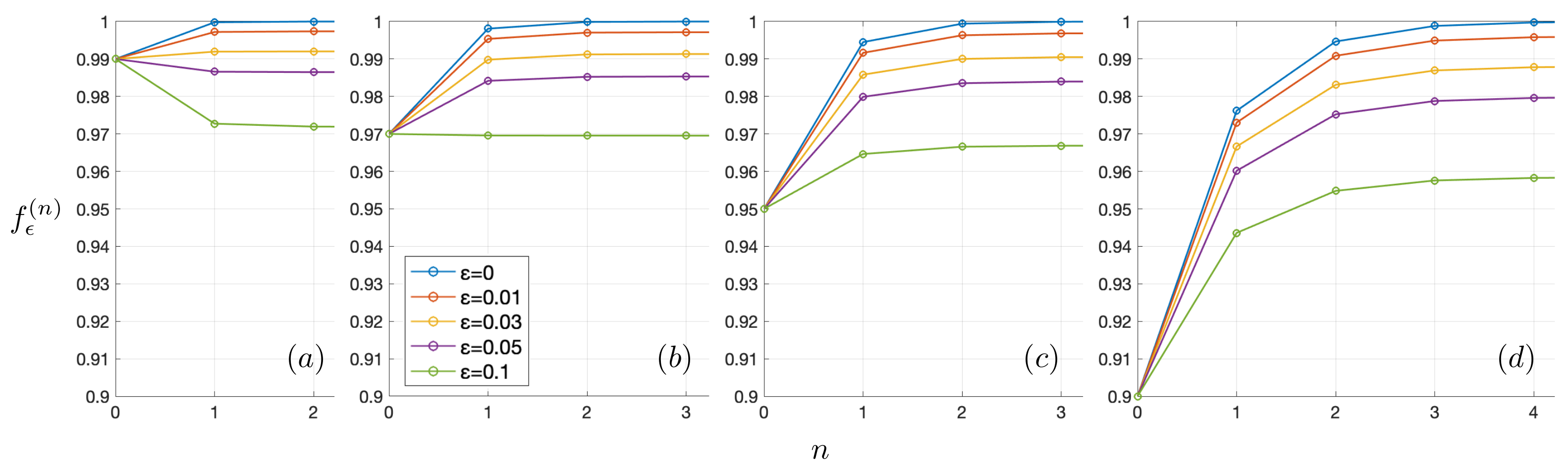}
    \caption{  The purification of noisy SPAM is demonstrated with a CNOT gate having an error rate $\epsilon$. The vertical axis is an initial fidelity of state preparation as the number of target qubits $n$ in the horizontal one increases. An initial fidelity and a noise fraction are balanced, i.e., $f=1-q$. (a) For an initial fidelity $f=0.99$,  a single target qubit suffices to purify a noisy state preparation. A noisy CNOT gate with an error rate $\epsilon > 0.0374$ cannot be used to increase the fidelity; see Table \ref{e}. (b) For $f=0.97$, the purification protocol immediately achieves noiseless state preparation with two target qubits. (c) For $f=0.95$, noiseless state preparation can be achieved with three target qubits. The purification protocol needs CNOT gates with an error rate of less than $0.146$. (d) For $f=0.9$, four qubits are needed for the protocol to reach noiseless measurements. For all cases, the purification protocol is efficient, and a few noisy target qubits are a resource to achieve noiseless SPAM. }
    \label{fig:graph}
\end{figure*}

\subsection*{Purification of noisy SPAM with noisy interactions}

So far, we have shown that purification works with the given resources of noisy SPAM for both cases of preparation and measurements. Namely, it suffices to work with noisy resources of SPAM {\it per se} to cleanse the errors from noisy SPAM. One can compare this to error correction, where noisy gates do not work to correct gate errors. Furthermore, quantum gates and SPAM working for diagnosing gate errors should contain sufficiently small error rates toward fault-tolerant computation \cite{doi:10.1137/S0097539799359385, 548464, doi:10.1126/science.279.5349.342}. 

We observe the fact that CNOT gates in the purification protocol are noiseless. For the full generality, we relax the assumption and consider the purification protocol. A noisy CNOT gate may be given as \cite{PhysRevLett.83.4200},
\bea
V_2 (~\cdot~) V_{2}^{\dagger}~~\mapsto ~~(1- \epsilon) V_2 (~\cdot~) V_{2}^{\dagger} +   \epsilon~\tr[~\cdot~] \frac{\I}{2} \otimes \frac{\I}{2},~~~~~ \label{eq:noisecnot}
\eea
where $\epsilon$ denotes a noise fraction. Then, noisy CNOT gates read the recurrence relation in the following while noisy state preparation gets purified, see also Eq. (\ref{eq:res}),
\begin{align}
R_{00}^{(n+1)} &= (1-\epsilon) \alpha R_{00}^{(n)}  + \frac{\epsilon}{4} (R_{00}^{(n)}+R_{11}^{(n)}),~\mathrm{and} \nonumber\\
R_{11}^{(n+1)} &= (1-\epsilon) (1-\alpha) R_{11}^{(n)}  + \frac{\epsilon}{4} (R_{00}^{(n)}+R_{11}^{(n)}). \nonumber
\end{align}
To see the convergence of a resulting state purified by $n$ qubits, let us compute the ratio,
\bea
g &=& \lim_{n\rightarrow \infty} \frac{R_{11}^{(n)}}{R_{00}^{(n)}} =  \sqrt{D^2 +1} -D, \nonumber \\
&& \mathrm{where} ~~ D = 2(2f -1)(1-2q)\frac{ (1-\epsilon)}{\epsilon}. \label{eq:d} 
\eea
Then, the purification with $n$ target qubits leads to a convergent state up to a fidelity in the following, 
\bea
f_{\epsilon}^{(n)} = \frac{R_{00}^{(n)} }{\tr R^{(n)}  } ~~  \xrightarrow{ n \rightarrow \infty }~~ \frac{1}{1-D+\sqrt{D^2 +1}}  ~~~~~\label{eq:fep}
\eea
where the convergence is strictly away from one, showing a limitation in the purification. As an instance, for realistic error rates $\epsilon=1-f=q=0.05$, the purification achieves a fidelity up to $0.984$, for which two target qubits suffice. That is, the protocol can purify noisy state preparation from $0.95$ up to $0.984$. The success probability with two target qubits is given by {$p_{succ}^{(3)}=0.7357$}. 

A similar conclusion may be drawn when noisy CNOT gates are applied to the purification of noisy measurements. The recurrence relation for a POVM element in Eq. (\ref{eq:rem}) is obtained,
\bea
\bracket{k}{\M_k^{(m+1)}}{k} & = & (1-\epsilon) \alpha \bracket{k}{\M_k^{(m  )}}{k}  + \frac{\epsilon}{4} \tr[\M_k^{(m )}], ~\mathrm{and}\nonumber \\
\bracket{\Bar{k}}{\M_k^{(m+1)}}{\Bar{k}} &=& (1-\epsilon)(1-\alpha) \bracket{\Bar{k}}{\M_k^{(m )}}{\Bar{k}}   + \frac{\epsilon}{4} \tr[\M_k^{(m )}]. \nonumber
\eea
Then, a noise fraction of a purified POVM element depends on both the number of target qubits $m$ and a noise fraction $\epsilon$; the noise fraction is also convergent,
\bea
q^{(m,\epsilon)} = \langle \bar{k} | \N_{k}^{(m,\epsilon)} | \bar{k} \rangle  ~~  \xrightarrow{ m \rightarrow \infty }~~ \frac{1}{1+D+\sqrt{D^2 +1}},   ~~~~~ \label{eq:mep}
\eea
which is strictly positive, thus not achieving a noiseless POVM element. One can notice that a noise fraction above is equal to $1-f_{\epsilon}^{(m)}$ in Eq. (\ref{eq:fep}). Then, as mentioned above, when a CNOT gate has an error rate $\epsilon =0.05$, the protocol suppresses an error rate from $0.05$ up to $0.016$. 

\subsection*{  The purification condition}  

From the result so far, it turns out that noiseless CNOT gates are crucial for the purification of noisy SPAM; it is now clear that with noisy CNOT gates, the purification protocol may enhance the fidelity and suppress error rates to some level, but eventually fails to achieve noiseless SPAM. As it is shown in Eqs. (\ref{eq:d}), (\ref{eq:fep}) and (\ref{eq:mep}), the convergence relies on noise parameters, $f$, $q$, and $\epsilon$. The purification protocol does not work if the first iteration fails to enhance the fidelity, which leads to the condition for purifying noisy SPAM;
\bea
\mathrm{the ~ purification~condition~}: f < f_{\epsilon}^{(1)}. \label{eq:pc}
\eea
As an instance, assuming that SPAM errors are balanced, i.e., $1-f=q$, an initial fidelity increases by the purification protocol whenever $\epsilon <  \epsilon_c$ where the threshold $\epsilon_c$ is given by,
\bea
 \epsilon_c & = &  \frac{  8 f^3 -12f^2 +4f }{ 8 f^3 -12 f^2 +4f -1} ~~~~~\label{eq:ec} \\
 & \approx & 4(1-f) - 28(1-f)^2 + O((1-f)^3). \nonumber
 \eea
Note that the expansion shows that when error rates $1-f=q$ are $10^{-2}$, a threshold of an error rate $\epsilon_c$ is about four times more relaxed. In Table \ref{e}, thresholds $\epsilon_c$ are computed for error rates relevant to NISQ technologies. For instance, when an error rate is $1-f=q=0.01$, the critical error rate $\epsilon_c$ is $0.0374$. CNOT gates noisier than given noisy SPAM can still be applied to cleanse the static errors. 

\begin{table}[h]
\centering
\begin{tabular}{ccccccccc}
\hline
\hline
~~$f$ ~ ~~\vline \vline && 1~~& 0.99 ~~&0.97~ & 0.95~ & 0.93~ & 0.9  \\  \hline
~~ $\epsilon_c$ ~ \vline\vline && 0~~& 0.0374 ~~ & 0.0986~~ & 0. 1460~~ & 0.1830~~ & 0.2236 \\\hline
 \hline
\end{tabular}
\caption{ Critical error rates $\epsilon_c$ of a CNOT gate below which noisy SPAM can be purified are shown. }\label{e}
\end{table}

\subsection*{  Verification of error rates }

The purification condition in Eq. (\ref{eq:pc}) tells that when a CNOT gate is noisy $\epsilon>0$, one need to verify error rates of SPAM $1-f$ and $q$ to find if a purification works, or not. We may compare to the case $\epsilon =0$ in which the verification of static errors is not necessary; the purification with a few qubits reaches almost a noiseless SPAM. For SPAM at a single-copy level, probabilities by a noisy measurement $\{ \M_0, \M_1\}$ for a noisy state $\rho$ show probabilities,
\bea
p(0) = f(1-q) + (1-f)q~~\mathrm{and}~ ~p(1)=1-p(0). \nonumber
\eea
Since the probabilities obtained from an experiment are not independent, they cannot single out the parameters $f$ and $q$, respectively. Without the verification of error rates, one cannot decide whether the purification protocol works to increase a fidelity of state preparation and suppress a measurement error.  

\begin{figure}[h!]
    \centering
    \includegraphics[width=0.43\textwidth]{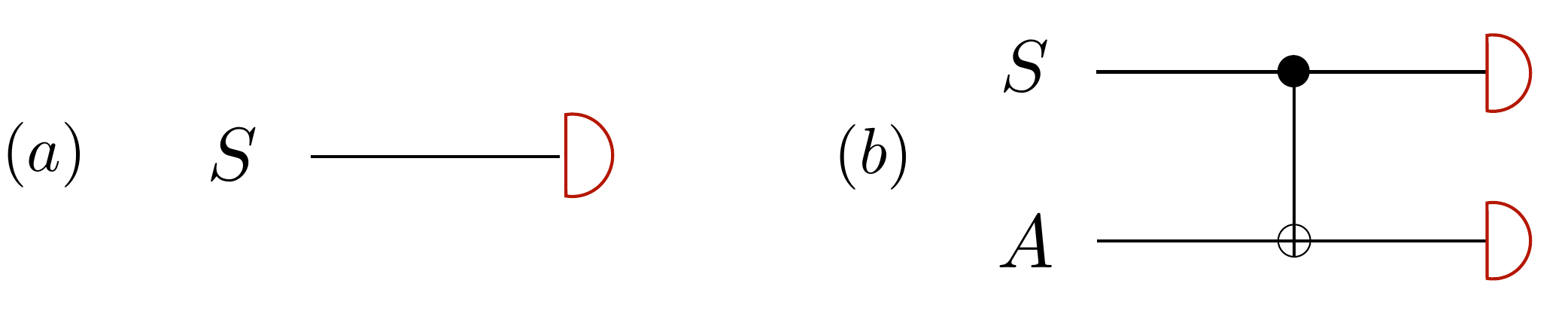}
    \caption{ (a) SPAM at a single-copy level cannot separate errors of preparation and measurements. (b) Error rates of SPAM and a CNOT gate, respectively, can be verified from measurement statistics, see Eq. (\ref{eq:v}).}
    \label{fig:spam}
\end{figure}

We resolve the verification of error rates by introducing a noisy CNOT gate, see Fig. \ref{fig:spam}, which allows to estimate error rates $1-f$, $q$, and $\epsilon$. Noisy measurements are performed after a noisy CNOT gate on noisy preparation. Probabilities $p(ij)$ for outcomes $i,j\in \{0,1 \}$ are determined by the error rates, 
\bea
p(ij) = F_{ij}(f,q,\epsilon),~~\mathrm{for}~~ i,j\in \{ 0,1 \} \label{eq:v}
\eea
for which the specific forms of $F_{ij}$ are shown in the following,
\begin{widetext}
\begin{align*}
p(00) & = (1-\epsilon)\big[f^2 (1-q)^2 + (1-f)^2 q^2 + f(1-f)(1-q)q  + (1-f)^2 q(1-q) + (1-f)fq^2\big] + \epsilon/4, \\
p(01) &= (1-\epsilon)\big[f^2 (1-q)q + (1-f)^2 q^2 + f(1-f)(1-q)^2  + (1-f)^2 q^2 + (1-f)fq(1-q)\big] + \epsilon/4, \\
p(10) &= (1-\epsilon)\big[f^2 q(1-q) + (1-f)^2 (1-q)q + f(1-f)q^2  + (1-f)^2 (1-q)^2 + (1-f)f(1-q)q\big] + \epsilon/4, \mathrm{and }\\
p(11) &= (1-\epsilon)\big[f^2 q^2 + (1-f)^2 (1-q)^2 + f(1-f) q(1-q)  + (1-f)^2 (1-q)q + (1-f)f(1-q)^2\big] + \epsilon/4.
\end{align*}  
\end{widetext}
Note that three of the equations are independent since $\sum_{ij} p(ij)=1$, from which the error rates $f$, $q$, and $\epsilon$ can be obtained. A closed form of the solution, i.e., the error rates in term of probabilities $p(ij)$, is not presented here; numerically it is fairly straightforward to compute the error rates with probabilities obtained experimentally.

 \subsection*{Demonstration of purification }  
 
Once error rates are verified, one can see if the purification in Eq. (\ref{eq:pc}) is fulfilled and also estimate the number of target qubits for the purification. In Table in the following, we demonstrate the verification of error rates and the purification, see also Fig. \ref{fig:graph}. 

In Table, Case 1 shows that the protocol purifies noisy SPAM when a CNOT gate is noiseless. An error rate of $10^{-3}$ is achieved with three additional target qubits. In Case 2, Case 3, and Case 4, a CNOT gate is noisy and noisy SPAM is purified up to $f^{(\infty)}$. It is shown that a few qubits are sufficient to achieve the purification. Case 5 shows an instance where the purification condition is not fulfilled; the error rate of a noisy CNOT gate is too high to increase the fidelity of noisy state preparation.

\begin{widetext}
\begin{tabularx}{0.95\textwidth} { 
  | >{\raggedright\arraybackslash} X 
  | >{\raggedright\arraybackslash}X 
  | >{\raggedright\arraybackslash}X  | }
 \hline
Measurement outcomes 

(p(01),p(10), p(11)) & 
Verified error rates

( 1-f, q, $\epsilon$ ) & 
Purified state preparation

with $n$ target qubits
\\
 \hline
 
 \hline
Case1: (0.154, 0.09, 0.09) 
& (0.1, 0.1, 0)  & 

 $f^{(1)}=0.976$ 

 $f^{(2)}=0.995$ 

 $f^{(3)}=0.999$ 

 $f^{(\infty)}=1$
 \\
 \hline
Case2: (0.1252, 0.0540, 0.0896) & (0.1, 0.05, 0.01)  & 
 
 $f^{(1)}=0.979$
 
 $f^{(2)}=0.994$ 
 
 $f^{(3)} = f^{(\infty)}=0.996$
 \\
 \hline
Case3: (0.0777, 0.0530, 0.0366) & (0.03, 0.05, 0.03)  & 

 
 $f^{(1)}=0.989$
 
 $f^{(2)}= f^{(\infty)}=0.991$
 \\
 \hline
 Case4: (0.0961, 0.0576, 0.0576) & (0.05, 0.05, 0.05) & 


$f^{(1)}=0.980$

$f^{(2)}=0.983$

$f^{(\infty)}=0.984$
\\
 \hline
Case5: (0.0754, 0.0674, 0.0357) & (0.01, 0.05, 0.1) &
 $f^{(0)}=0.99$
 
 $f^{(1)}=0.971$
 
 $f^{(\infty)}=0.970$
 \\




  
\hline
\end{tabularx}
\end{widetext}
 

It is shown that in all cases the protocol is efficient and few noisy target qubits are immediately a cost-effective resource for the purification.

\begin{figure}[h]
    \centering
    \includegraphics[width=0.47\textwidth]{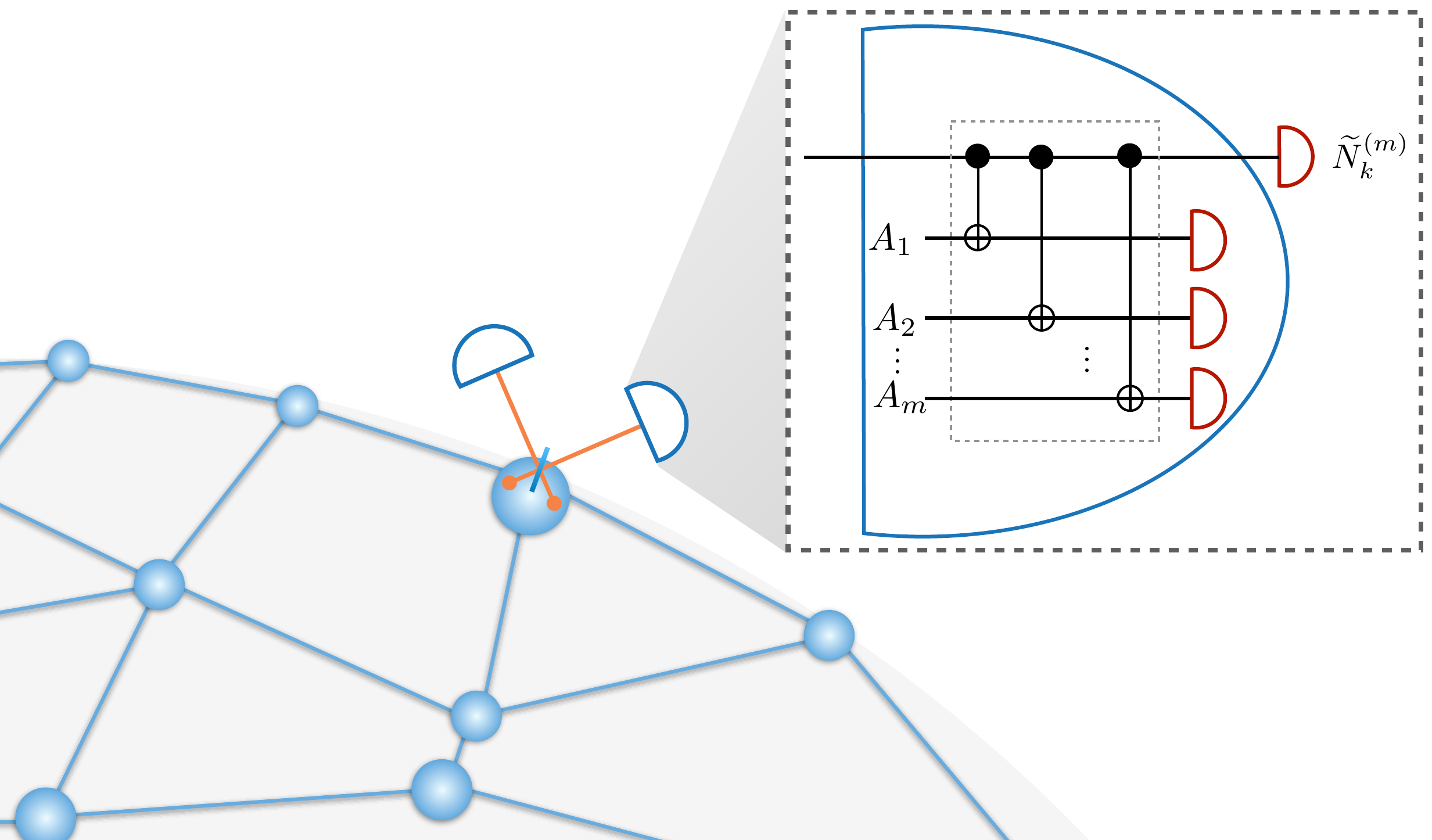}
    \caption{A quantum repeater realizes entanglement swapping with a Bell measurement that can be decomposed into a CNOT and single-qubit gates and local measurements. A network can establish noiseless entanglement with purified measurements.    
     }
  \label{fig:network}
\end{figure}

 \subsection*{ Applications to quantum communication }  

A measurement setting appears in quantum communication scenarios in various ways and our results presents a versatile tool to tame noisy preparation and noisy measurements. On the one hand, a measurement describes detection events on quantum systems either sent through a channel or entangled with other systems. For instance, fundamental tasks such as quantum state discrimination \cite{Barnett:09, Bergou:2007aa, Bae_2015}, quantum steering \cite{ PhysRevLett.98.140402,Cavalcanti:2017aa, RevModPhys.92.015001}, Bell tests \cite{RevModPhys.86.419}, and quantum pseudo-telepathy \cite{Brassard:2003aa}, which are closely connected or applied to quantum cryptographic protocols or communication primitives, e.g., quantum random access codes \cite{10.1145/581771.581773}. In these cases, communication protocols with noisy measurements may not realize the quantum advantages in communication. Hence, purifying noisy measurements may be a pathway to retain the advantages in two-party communication tasks. 

On the other hand, a quantum network consisting of both quantum and classical nodes involves repeaters that aim to realize entanglement swapping toward entanglement distribution over multiple parties, see Fig. \ref{fig:network}. Noisy measurements in entanglement swapping lead to noisy entanglement that may end up with undistillable states such as bound entangled states or even separable ones \cite{Chruscinski:2014aa}. The purification of noisy measurements may suppress error rates to enable a network to establish noiseless and large-size entanglement. 


 \subsection*{ Concluding remarks }  

We have characterized static quantum errors and presented a protocol for purifying noisy SPAM to increase the fidelity of state preparation and suppress an error rate in measurements. The purification of noisy SPAM is a versatile tool to tame noisy quantum resources for preparation and measurements. It only exploits noisy SPAM to cleanse noisy SPAM via a collection of CNOT gates. The protocol is efficient such that a few noisy qubits are immediately cost-effective for cleaning noise on a system qubit, up to $10^{-3}$, both state preparation and measurement. We have also shown how to estimate error rates in SPAM with probabilities obtained from experiments and, with verified error rates, demonstrated the purification. 

Our findings have significant implications to quantum information applications in the era of NISQ technologies. Purifying noisy SPAM will improve quantum algorithms relying on NISQ technologies, where SPAM is repeatedly performed without a cost between iterations of short-depth quantum circuits \cite{RevModPhys.94.015004, TILLY20221}. It may also be used to enhance quantum measurements for communication purposes in two-party protocols as well as a network scenario. With purified measurements, quantum advantages in communication would be attained and a quantum network may share noiseless entanglement over an arbitrary distance. Our results are readily feasible with currently available quantum technologies.

The purification of measurements can also be exploited in fundamental experiments in quantum information, particularly quantum certifications. For instance, a sufficiently high noise fraction on incompatible measurements \cite{PhysRevLett.122.130402}, equivalently quantum steering \cite{PhysRevLett.115.230402}, results in compatible ones. In addition, noisy measurements constituting entanglement witnesses for verifying entanglement may fail to detect entangled states; noisy observables may have positive expectation values for all quantum states \cite{GUHNE20091, Chruscinski:2014aa}. Purifying measurements help the certification tasks, which we envisage will also enable quantum information applications relying on the fundamental certifications.

\subsection*{Acknowledgement}

This work is supported by the National Research Foundation of Korea (Grant No. NRF-2021R1A2C2006309, NRF-2022M1A3C2069728, RS-2024-00408613, RS-2023-00257994) and the Institute for Information \& Communication Technology Promotion (IITP) (RS-2023-00229524). 


%

\end{document}